\documentclass{SciPost}

\hypersetup{
    colorlinks,
    linkcolor={red!50!black},
    citecolor={blue!50!black},
    urlcolor={blue!80!black}
}

\usepackage[bitstream-charter]{mathdesign}
\usepackage{float}
\usepackage{bm}
\usepackage{soul}

\DeclareSymbolFont{usualmathcal}{OMS}{cmsy}{m}{n} 
\DeclareSymbolFontAlphabet{\mathcal}{usualmathcal}

\fancypagestyle{SPstyle}{
\fancyhf{}
\lhead{\colorbox{scipostblue}{\bf \color{white} ~SciPost Physics }}
\rhead{{\bf \color{scipostdeepblue} ~Submission }}

\fancyfoot[C]{\textbf{\thepage}}
}

\begin{document}

\pagestyle{SPstyle}

\begin{center}{\Large \textbf{\color{scipostdeepblue}{
BBP transition and the leading eigenvector of the spiked Wigner model with inhomogeneous noise
}}}\end{center}

\begin{center}\textbf{
Leonardo S. Ferreira \textsuperscript{$\star$} and
Fernando L. Metz\textsuperscript{$\circ$}
}\end{center}

\begin{center}
Physics Institute, Federal University of Rio Grande do Sul, 91501-970 Porto Alegre, Brazil
\\[\baselineskip]
$\star$ \href{mailto:email1}{\small leop.dsf@gmail.com},
$\circ$  \href{mailto:email4}{\small fmetzfmetz@gmail.com}
\end{center}

\section*{\color{scipostdeepblue}{Abstract}}
\boldmath\textbf{The spiked Wigner ensemble is a prototypical model for high-dimensional inference. We
  study the spectral properties of an inhomogeneous rank-one spiked Wigner model in
  which the variance of each entry of the noise matrix is itself a random variable. In the high-dimensional limit, we derive exact equations
  for the spectral edges, the outlier eigenvalue, and the distribution of the components of the outlier eigenvector. These equations determine
  the BBP transition line that separates the gapped phase, where the signal is detectable, from the gapless phase.
  In the gapped regime, the distribution of the outlier eigenvector provides
  a natural estimator of the spike. We solve the equations for a noise matrix whose variances
  are generated from a truncated power-law distribution.
  In this case, the BBP transition line is non-monotonic, showing that
  an inhomogeneous noise can enhance signal detectability.
}

\vspace{\baselineskip}

\noindent\textcolor{white!90!black}{%
\fbox{\parbox{0.975\linewidth}{%
\textcolor{white!40!black}{\begin{tabular}{lr}%
  \begin{minipage}{0.6\textwidth}%
    {\small Copyright attribution to authors. \newline
    This work is a submission to SciPost Physics. \newline
    License information to appear upon publication. \newline
    Publication information to appear upon publication.}
  \end{minipage} & \begin{minipage}{0.4\textwidth}
    {\small Received Date \newline Accepted Date \newline Published Date}%
  \end{minipage}
\end{tabular}}
}}
}


\vspace{10pt}
\noindent\rule{\textwidth}{1pt}
\tableofcontents
\noindent\rule{\textwidth}{1pt}
\vspace{10pt}


\section{Introduction}

Spiked matrix ensembles are paradigmatic models in high-dimensional statistics \cite{Miolane2019}. First introduced 
in the context of probabilistic principal
component analysis \cite{Johnstone2001}, they provide fundamental insights into a broad class of inference problems, including graph
clustering \cite{Abbe2018}, synchronization on compact groups \cite{Perry-Sync2018}, and community detection \cite{Abbe2018}.
In signal processing, spiked matrix ensembles model an observation matrix through a signal-plus-noise decomposition, providing
an ideal setting for studying the inference of a rank-one signal 
corrupted by additive noise. The central question is then to determine under which conditions the signal can
be detected and recovered in the high-dimensional limit \cite{Miolane2019}.

Random matrix theory plays a central role in this class of problems \cite{Peche2006,Tao2013,BBP,Barbier2022,Barbier2025}, since the observation
matrix takes the form of a low-rank perturbation of a large random matrix.
A prototypical model in this context is the spiked Wigner
ensemble \cite{Miolane2019,Peche2006,Maida2007,Feral2007,Knowles2013}, in which the entries of the noise matrix
are independent and identically distributed random variables.
As the signal strength increases, this model undergoes a well-known spectral transition \cite{Peche2006,Tao2013}: above a critical
threshold, a single eigenvalue detaches from the bulk of the noise spectrum \cite{Wigner51}, and the eigenvector associated with this
outlier eigenvalue becomes correlated with the spike vector, thereby providing a spectral estimator of the signal \cite{B-GN}.
This threshold phenomenon is commonly referred to as
the Baik-Ben Arous-Péché (BBP) transition \cite{BBP}.

The performance of spectral inference methods depends strongly on the assumptions
made on the observation matrix. In particular, the effect of structural properties of the rank-one
signal has been widely studied. Sparsity, for
instance, has been incorporated into the signal to account for sample
incompleteness \cite{Candes2006,Deshpande2014,Lesieur2015}, while correlations between signal
components have been introduced through generative models \cite{GenerativeBruno}.
From the viewpoint of random matrix theory, however, modifying the structure of the rank-one perturbation
mainly affects the outlier eigenvalue and its associated eigenvector, while the bulk of
the spectrum remains controlled by the noise matrix \cite{Tao2013}. It is therefore natural to ask how
structural features of the noise matrix shape the spectral properties of the ensemble.
In the case of sparse noise matrices, for example, structural fluctuations may generate extremely large
eigenvalues \cite{Krivelevich2003,Chung2003}, producing a tail in the spectrum that can suppress the gapped
regime associated with the BBP transition and limit the applicability of spectral inference methods.
Furthermore, applications in bioinformatics \cite{GWassociation} and recommendation
systems \cite{Netflix} suggest that the noise matrix may also contain a nonrandom component
arising from structural features not captured by the signal alone. These observations
highlight the importance of incorporating
structure directly into the noise matrix.

In fact, recent works \cite{Barbier2025,Barbier23,Guionnet2025,Adomaityte25} have introduced
structured versions of the spiked Wigner model in which the standard assumptions of independence
and identical distribution are relaxed.
One line of work \cite{Barbier2025,Barbier23} focuses on relaxing independence by
considering rotationally invariant noise matrices with arbitrary potentials, thus introducing
correlations among the matrix entries. Another direction, developed in \cite{Guionnet2025}, relaxes the assumption
of identical distribution by considering inhomogeneous noise matrices, including a specific model with variances arranged
in a block structure. This setting is closely related to multi-species spin-glass models \cite{MultiSpin,Panchenko2015} and inference models with spatial
coupling \cite{Javanmard2013,Barbier2016}.
By mapping the block-structured ensemble to an effective homogeneous form, the signal-to-noise threshold for the BBP transition has been
determined in terms of the variance profile \cite{Guionnet2025}.

 A distinct class of structured models has been considered in \cite{Adomaityte25}, where the noise matrix is sparse and its
 entries encode the edges of a random graph. Although the spectral properties of sparse random matrices have been extensively
 studied using replica and cavity methods \cite{EdwardsJones1976,MNR2019}, these approaches typically yield a set of distributional equations for the
 resolvent whose solution generally relies on numerical methods. An alternative route to obtaining nontrivial analytic solutions was
 introduced in \cite{Metz2020,Jeferson2022}. By considering random graphs with an arbitrary degree distribution, these
 works solved the resolvent equations in the limit of an infinitely large mean degree. Remarkably, in this so-called
 high-connectivity limit, random graphs retain signatures of their heterogeneous structure, giving rise to a novel
 class of effective fully connected ensembles whose spectral properties depend explicitly on the degree
 distribution. This effective description of the adjacency matrix of highly connected random graphs provides a
 natural framework for incorporating structural heterogeneity into the noise matrix of spiked models.

Motivated by these recent developments, we study an inhomogeneous spiked Wigner model
in which the entries of the noise matrix are not identically distributed. In contrast
to reference \cite{Guionnet2025}, where the variance profile is deterministic, we consider
an ensemble in which the variances of the matrix elements are themselves random
variables. This extension allows us to
investigate how disorder in the variance profile modifies the spectral
properties of the model and, in particular, how it affects the existence of a gapped phase and the
detectability of the signal by spectral methods. The proposed ensemble for the
  noise matrix also represents the high-connectivity
limit of the adjacency matrix of random graphs with arbitrary degree
distributions \cite{Metz2020,Dembo2021,Jeferson2022}.

We derive exact equations for the spectral properties of the observation matrix and use them to determine
when spectral methods can detect and estimate the signal in the high-dimensional limit.
Specifically, we obtain equations for
the spectral edges of the bulk and for the outlier eigenvalue, whose solutions yield
the BBP critical line as a function of the variance profile, extending the class of spiked
matrix ensembles for which sharp detectability thresholds can be established. 
When an outlier eigenvalue is present, the associated eigenvector
provides a natural estimator of the spike vector. While the location of outlier eigenvalues has been extensively
studied in this context \cite{Peche2006,Maida2007,Feral2007,Knowles2013,Tao2013}, much less is known about the eigenvectors associated
with extreme eigenvalues. Previous works have mainly focused on the overlap between the
outlier eigenvector and the signal in spiked Wigner \cite{B-GN} and
Wishart \cite{Hoyle2007,Paul2007,BouchaudCleaning} ensembles. Here,  we derive
an analytic expression for the full distribution of the components of the outlier eigenvector
as a function of the variance profile and the spike prior, which yields a simple
prescription to estimate the signal and compute the associated mean-squared error.
We illustrate our theory in a model where the individual
variances follow a truncated power-law distribution. Interestingly, the BBP critical line of this model exhibits
a non-monotonic behaviour as a function of the model parameters, showing that variance fluctuations
can enhance detectability.
Our theoretical findings are in excellent agreement with numerical diagonalizations of finite-dimensional observation matrices.

The remainder of the paper is organized as follows. In section \ref{sec-2}, we define the inhomogeneous spiked Wigner model, review the main
spectral properties of the homogeneous case, and discuss the relation between our model
and previous works. 
In section \ref{derivation}, we derive the main equations for the spectral properties of the observation matrix for
arbitrary spike priors and variance profiles. Section \ref{power-law} presents the results obtained by solving these
equations for a truncated power-law distribution of the variables defining the variance profile. Finally, section \ref{conclusion} summarizes the
main results and presents our conclusions, while the appendix explains the derivation of the distributional equations that form
the core of our analytic results.


\section{The inhomogeneous spiked Wigner model} \label{sec-2}

We address the problem of estimating a signal vector from observations of a matrix with an inhomogeneous noise structure. The data
is modeled by a real symmetric random matrix  $\boldsymbol{A}_n \in\mathbb{R}^{n\times n}$ of the form
\begin{equation}\label{model}
    \boldsymbol{A}_n=\frac{1}{\sqrt{n}}\left(\boldsymbol{S}_{n}^{\odot 1/2} \odot \boldsymbol{W}_n \right) + \frac{\gamma}{n}\boldsymbol{X}_n\boldsymbol{X}_n^T,
\end{equation}
where $\odot$ is the Hadamard product and $(\dots)^T$ denotes the transpose. The signal $\boldsymbol{X}_n \in\mathbb{R}^n$ is formed by i.i.d.
components drawn from a distribution $P_{X}(x)$ with mean $\langle X \rangle > 0$ and second moment $\langle X^2 \rangle$.
The parameter $\gamma \geq 0$ is the signal-to-noise ratio that quantifies the signal strength. The noise component in Eq. (\ref{model}) is defined
in terms of two symmetric matrices: $\boldsymbol{W}_n \in\mathbb{R}^{n\times n}$ and $\boldsymbol{S}_n \in\mathbb{R}^{n\times n}$. The entries of $\boldsymbol{W}_n$ are i.i.d.
random variables sampled from a distribution with zero mean and unitary variance. The elements of $\boldsymbol{S}_n$ are given by $S_{ij}=S_iS_j$, where
$\{ S_i \}_{i=1}^n$ are i.i.d. variables drawn from a probability density $P_S(s)$ with a finite support $\Omega_S$ on the non-negative real line. Since
$S_i S_j/n$ is the variance of the $ij$-th entry of the noise matrix, we refer to $P_S(s)$ as the {\it variance profile}.

The noise component in Eq. (\ref{model}) naturally arises from the high-connectivity limit of heterogeneous
  random graphs \cite{Metz2020,Jeferson2022}. Let $H_{ij} = C_{ij} J_{ij}$ denote the entries of the weighted adjacency matrix of a random
  graph, where $\boldsymbol{C}_n \in\{0,1\}^{n\times n}$ is the adjacency matrix and $\boldsymbol{J}_n \in\mathbb{R}^{n\times n}$ is the matrix
  of coupling strengths, with elements $J_{ij} \in \mathbb{R}$. The rescaled degree of node $i$ is defined by $K_i=c^{-1}\sum_{j=1}^n C_{ij}$, where $c$ is the mean
  degree. We assume that $\{ K_i \}_{i=1}^N$ are i.i.d. random variables drawn from an arbitrary distribution $P_{\rm resc}(K)$, while
  $\{ J_{ij} \}_{ij = 1}^n$ are i.i.d. random variables with zero mean and variance $1/c$.
  Taking first the thermodynamic limit $n \rightarrow \infty$ and then the high-connectivity limit $c \rightarrow \infty$, the spectral
  density of $\boldsymbol{H}_n$ converges to the free multiplicative convolution of $P_{\rm resc}(K)$ with the Wigner
  semicircle law \cite{Metz2020,Dembo2021}. This result implies the following  identity for the weighted adjacency matrix in the limit $c \rightarrow \infty$ \cite{Jeferson2022}
\begin{equation}
\boldsymbol{H}_n
= \frac{1}{\sqrt{n}} \boldsymbol{K}_{n}^{\odot1/2}\boldsymbol{W}_{n}\boldsymbol{K}_{n}^{\odot1/2},
\label{juop}
\end{equation}
where $\boldsymbol{K}_n$ is a diagonal random matrix with entries $K_{ij} = K_i \delta_{ij}$.
The exact correspondance between $\boldsymbol{A}_n$ and $\boldsymbol{H}_n$ provides a natural
interpretation of the inhomogeneous model.
Equation (\ref{model}) therefore describes a spiked random matrix
whose noise component is generated by the high-connectivity limit of heterogeneous random graphs.

The matrix $\boldsymbol{A}_n$ models the observations of a rank-one signal $\frac{\gamma}{n}\boldsymbol{X}_n  \boldsymbol{X}_{n}^T$ corrupted by an additive
noise matrix $\frac{1}{\sqrt{n}}(\boldsymbol{S}_{n}^{\odot 1/2} \odot \boldsymbol{W}_n)$.
The scaling with the matrix dimension $n$ in Eq. (\ref{model}) ensures that the eigenvalues of $\boldsymbol{A}_n$ remain of $\mathcal{O}(n^0)$
as $n \rightarrow \infty$. The fluctuations
of $\{ S_i \}_{i=1}^n$ control the inhomogeneity of the noise.
In the limit $n \rightarrow \infty$, the spectrum of $\boldsymbol{A}_n$
is generally dominated by the continuous part of the spectrum, which does not contain
meaningful information about the rank-one signal matrix. The questions we will address in this work concern the role played by the
variance profile $P_S(s)$ or the inhomogeneous structure
of the noise on the detection and recovery of $\boldsymbol{X}_n$ \cite{Perry2018}:
\begin{itemize}
\item[$\bullet$] Detection: can the spectrum of the observation matrix $\boldsymbol{A}_n$ be distinguished from
  the spectrum of the pure noise component? How does the variance profile $P_S(s)$ influence
  the detection of the signal?
\item[$\bullet$] Recovery: once there is a clear distinction between signal and noise in the spectrum
  of $\boldsymbol{A}_n$, can we compute an eigenvector estimator
  for the spike vector? How
  does the variance profile $P_S(s)$ affect the quality of the signal recovery? 
\end{itemize}
We will derive results for the spectral properties of $\boldsymbol{A}_n$ in light of the above two aspects of the problem.


\subsection{Spectral properties of interest}

Let $\{ \lambda_i(\boldsymbol{A}_n)  \}_{i=1}^n$ be the eigenvalues of the observation matrix $\boldsymbol{A}_n$.
The primary quantity of interest is the empirical spectral density  
\begin{equation}\label{esd-def}
\rho(\lambda)=\lim_{n\to\infty}\frac{1}{n}\sum_{i=1}^n \delta\left[ \lambda-\lambda_i(\boldsymbol{A}_n) \right].
\end{equation}
The variance profile $P_S(s)$ has a bounded support, implying
that $\rho(\lambda)$ is nonzero on a finite interval $\Omega_{\lambda}= (-\lambda_{*},\lambda_{*})$. The quantity $\lambda_* > 0$ defines the spectral edge of $\rho(\lambda)$.
In the limit $n \rightarrow \infty$, the spectra of rank-one perturbations of large random matrices
typically contain a continuous part, forming the so-called bulk of eigenvalues, and eventually a single outlier eigenvalue $\lambda_{\rm out}$ \cite{Peche2006,Tao2013,O2016}, separated
from the bulk by a gap $\lambda_{\rm out} - \lambda_{*} >0$.  When the largest eigenvalue of the observation matrix $\boldsymbol{A}_n$
is equal to $\lambda_{\rm out}$, detection of the vector signal $\boldsymbol{X}_n$ is possible, and $\boldsymbol{X}_n$ can be estimated from
the components of the outlier eigenvector \cite{B-GN}. 
An eigenvector $\boldsymbol{R}_n(\lambda) \in \mathbb{R}^n$
associated to an eigenvalue $\lambda(\boldsymbol{A}_n)$ satisfies $\boldsymbol{A}_n \boldsymbol{R}_n(\lambda) = \lambda(\boldsymbol{A}_n) \boldsymbol{R}_n(\lambda)$, while
the empirical distribution of the eigenvector components $\{ R_{n,i}(\lambda) \}_{i=1}^n$ is defined as
\begin{equation}
P_R(r|\lambda) = \lim_{n \rightarrow \infty} \frac{1}{n} \sum_{i=1}^n \delta\left[ r - R_{n,i}(\lambda) \right].
\end{equation}  
Our goal is to derive equations for the spectral density $\rho(\lambda)$, the spectral edge $\lambda_*$, the
outlier $\lambda_{\rm out}$, and the eigenvector distributions $P_R(r|\lambda_{\rm out})$ and  $P_R(r|\pm \lambda_{*})$,  for arbitrary $P_X$ and $P_S$, which
will allow us to fully characterize the detection and recovery of $\boldsymbol{X}_n$ through spectral methods.


\subsection{Relation to previous work}

\begin{figure}[h!]
    \centering
    \includegraphics[width=8.5cm]{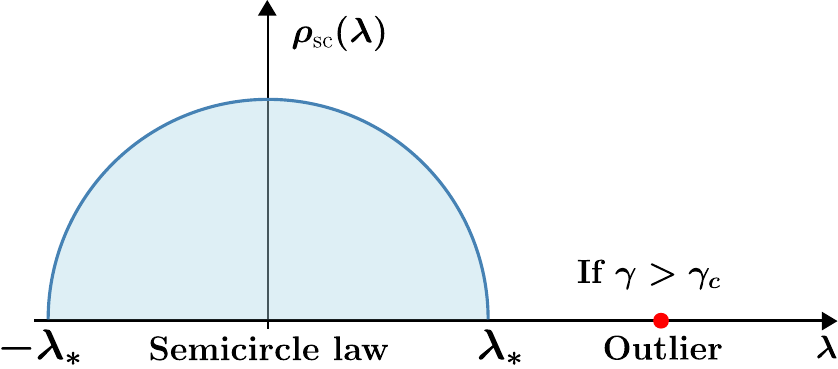}
    \caption{Spectrum of the homogeneous spiked Wigner model with $\sigma=1$ (see the main text). For $\gamma > \gamma_c$, an outlier eigenvalue (red dot) detaches from the bulk.
      For $\gamma \leq \gamma_c$, the largest eigenvalue is given by the upper spectral edge $\lambda_* = 2 \sigma$.}
    \label{fig:1}
\end{figure}

Equation (\ref{model}) generalizes the homogeneous spiked Wigner model \cite{Peche2006} to the case in which
the entries of the noise matrix are not identically distributed.
The homogeneous model is recovered by setting $P_S(s)=\delta(s-\sigma)$, with $\sigma > 0$.
In this case, the elements of the noise matrix in Eq. (\ref{model}) have constant variance $\sigma^2/n$, and the
spectral density follows the Wigner semicircle law \cite{Wigner51}
\begin{equation}\label{wigner}
    \rho_{\rm sc}(\lambda)= \frac{1}{2\pi \sigma^2} \sqrt{4\sigma^2-\lambda^2}, 
\end{equation}
supported on the interval $\Omega_{\lambda} = (-2\sigma,2\sigma)$.

The homogeneous spiked Wigner model undergoes a BBP transition at the critical line \cite{Edwards1976}
\begin{equation}\label{snr-wigner}
    \gamma = \sigma/\langle X^2\rangle,
\end{equation}
where $\langle X^2\rangle$ is the second moment of $P_X$.
When $\gamma > \sigma/\langle X^2\rangle$, the signal is strong enough for the spectrum of $\boldsymbol{A}_n$ to be distinguishable
from that of the noise matrix. In this regime, the spectrum contains an outlier \cite{Edwards1976}
\begin{equation}\label{out-wigner}
    \lambda_{{\rm out}} = \gamma \langle X^2\rangle+\frac{\sigma^2}{\gamma\langle X^2\rangle}
\end{equation}
separated from the bulk by a gap $\lambda_{\rm out}-\lambda_* > 0$, with $\lambda_{*} = 2 \sigma$. In this gapped phase, signal
detection is possible, since the outlier eigenpair carries information about $\boldsymbol{X}_n$. More
specifically, the eigenvector $\boldsymbol{R}_n(\lambda_{\rm out})$ associated
to $\lambda_{\rm out}$ is positively correlated with $\boldsymbol{X}_n$. 
When $\gamma \leq \sigma/\langle X^2\rangle$, there is no outlier and the spectrum
of $\boldsymbol{A}_n$ is gapless.
In this regime, detection of $\boldsymbol{X}_n$ by spectral methods is not possible, since the correlation between $\boldsymbol{X}_n$ and the eigenvector
at the spectral edge vanishes. A schematic illustration of the spectrum of the homogeneous spiked Wigner model is shown in figure \ref{fig:1}.

Recently, random-matrix models with structured noise have attracted
considerable attention \cite{Barbier2025,Barbier23,Guionnet2025}.
In particular, reference \cite{Guionnet2025} considers, for instance, a noise matrix whose variances $\{ S_{ij} \}_{i,j=1}^N$
are organized in a finite number of blocks, so that entries within the same block are identically distributed.
By applying a suitable rescaling of Eq. (\ref{model}), the authors map the structured noise ensemble back to a homogeneous form, which
allows them to determine the critical signal-to-noise ratio of
the BBP transition in terms of the variance profile.
While Eq. (\ref{model}) is closely related to the model of \cite{Guionnet2025}, we consider
a different class of variance matrices $\boldsymbol{S}_n$. Rather than assuming a deterministic variance profile, we
set $S_{ij}=S_iS_j$, where $\{ S_i \}_{i=1}^N$ are i.i.d. random variables drawn from $P_S$. The distribution $P_S$ shapes
the structure of the noise matrix and thus controls the inhomogeneity of the model. As we show below, this factorized
form of $S_{ij}$ allows us to make significant progress in characterizing the spectral properties of Eq. (\ref{model}) in
the limit $n \to \infty$. In particular, we derive exact  equations for the spectral density
of the observation matrix, the spectral edges, the BBP transition line, the outlier eigenvalue, and the distribution of
the corresponding eigenvector, for arbitrary distributions $P_X$ and $P_S$. These equations provide a detailed understanding of how
inhomogeneous noise affects spectral inference methods.


\section{Analytic results for the spectral properties}
\label{derivation}

In this section, we present the main analytic results for the spectral properties of $\boldsymbol{A}_n$ that are relevant to characterize the detection and recovery
of the rank-one signal. First, we provide equations that determine the continuous spectral density and
its support. We also derive an equation for the average outlier eigenvalue and the full
distribution of its associated eigenvector components, which yield a natural estimate of the rank-one signal in the gapped phase.
Finally, we obtain an analytic expression for the mean squared error between the outlier eigenvector and the rank-one signal. This quantity characterizes
the quality of the eigenvector estimator.
All results in this section hold for arbitrary distributions $P_X$ and $P_S$.

The spectral properties of random matrices are encoded in the $n \times n$ resolvent matrix
\begin{equation}
  \boldsymbol{G}_n(z)=(z\boldsymbol{I}_n-\boldsymbol{A}_n)^{-1},
    \label{res}
\end{equation}  
where $\boldsymbol{I}_n$ is the $n \times n$ identity matrix and $z=\lambda-i\epsilon$, with $\epsilon > 0$ and $\lambda \in \mathbb{R}$. 
The spectral density in Eq. (\ref{esd-def}) can be written in terms of the diagonal elements of $\boldsymbol{G}_n(z)$ as \cite{Mirlin2000}
\begin{equation}\label{ESD-sum}
    \rho(\lambda) = \lim_{\epsilon \to 0^+}\lim_{n\to\infty} \frac{1}{n\pi} \sum_{i=1}^n {\rm Im} \, G_{n,ii}(\lambda-i\epsilon).
\end{equation}
In the limit $\epsilon \rightarrow 0^{+}$, the diagonal components $\{ G_{n,ii}(z) \}_{i=1}^n$ have non-zero imaginary
parts inside the bulk of eigenvalues ($\lambda \in \Omega_{\lambda}$), being purely real outside the
bulk ($\lambda \notin \Omega_{\lambda}$). In addition, the resolvent elements satisfy $\lim_{|z|\to\infty} G_{n,ii}(z) = 0$. As shown in a series
of works \cite{MNR2019,NeriMetz2016,Neri2020,Metz2021}, the diagonal entries of $\boldsymbol{G}_n(z)$
also give access to the statistics of the eigenvector components outside the bulk. In fact, by ordering
the eigenvalues of $\boldsymbol{A}_n$ as $\lambda_1(\boldsymbol{A}_n) \geq \lambda_2(\boldsymbol{A}_n) \geq \dots \geq \lambda_n(\boldsymbol{A}_n)$, the $i$-th
component of the leading eigenvector $\boldsymbol{R}_{n}(\lambda_1)$ follows from the identity
\begin{equation}\label{r-def}
   \left(\sum_{k=1}^n R_{n,k}(\lambda_1)  \right) R_{n,i}(\lambda_1)   = -i\lim_{\epsilon \to 0^+} \epsilon\left[ \boldsymbol{G}_n(\lambda_1 - i\epsilon) \boldsymbol{1}_n \right]_i,
\end{equation}
where $\boldsymbol{1}_n=(1,\cdots,1)^T$. Explicit derivations of Eq. (\ref{r-def}) appear in  \cite{MNR2019,NeriMetz2016}.
The above identity enables to compute the eigenvector distribution $P_R(r|\lambda)$ for $\lambda=\lambda_{*}$ and $\lambda = \lambda_{\rm out}$ \cite{Metz2021}.
Thus, we have recast the problem of obtaining the spectral
properties of $\boldsymbol{A}_n$ into the computation of the resolvent matrix $\boldsymbol{G}_n(z)$.


\subsection{Inside the bulk: the empirical spectral density}

The first step is to determine the elements $\{ G_{n,ii}(z) \}_{i=1}^{n}$, which yield the spectral density through Eq. (\ref{esd-def}). By
using the cavity method for dense random matrices \cite{Cizeau1994}, we show in appendix \ref{appecav} that, in
the limit $n \rightarrow \infty$, the diagonal elements of the resolvent satisfy the distributional equation
\begin{equation}
  G(z) \stackrel{\rm d}{=} \frac{1}{z - q(z) S},
  \label{juhrt}
\end{equation}  
where $q(z) = \left\langle S \mkern1mu G(z) \right\rangle$ is the expected value of the product $S \mkern1mu G(z)$, with $S$ distributed as
$P_S$ and $G(z)$ distributed according to Eq. (\ref{juhrt}).
Here and elsewhere, the brackets $\langle F(Y_{1},\dots, Y_{M} ) \rangle$ represent the ensemble average
of an arbitrary function $F(Y_{1},\dots, Y_{M} )$ of $M$ random variables $Y_{1},\dots, Y_{M}$ with their corresponding joint distribution. Although $G_{ii}$ and $S_i$ are not
independent, it is straightforward to obtain from Eq. (\ref{juhrt}) a self-consistent equation for $q(z)$,
\begin{equation}\label{q-final}
    q(z) = \int_{\Omega_S} \frac{ds \,s P_S(s)}{z-q(z)s}.
\end{equation}
The solutions of Eqs. (\ref{juhrt}) and (\ref{q-final}) give the joint distribution of the real and imaginary parts of the diagonal elements of the resolvent
in the complex plane \cite{Jeferson2022}. From Eqs. (\ref{ESD-sum}) and (\ref{juhrt}), we get the following expression for the spectral density 
\begin{equation}\label{ESD-final}
    \rho(\lambda)=\frac{1}{\pi}\lim_{\epsilon \to 0^+} {\rm Im} \left[ \int_{\Omega_S} \frac{ds \, P_S(s)}{z-q(z) s} \right].
\end{equation}
Equations (\ref{q-final}) and (\ref{ESD-final}) have been originally derived in references \cite{Metz2020,Jeferson2022} by taking the high-connectivity limit of the spectral
density of random graphs with arbitrary degree distributions. The term in brackets in Eq. (\ref{ESD-final}) represents
  the ensemble average $\langle G(z)\rangle$ of the diagonal elements of the resolvent. When Eq. (\ref{q-final}) admits
  multiple solutions, the correct one is selected from the conditions $\text{Im}\langle G(z)\rangle \geq 0$ and $\lim_{|z| \rightarrow 0}\langle G(z)\rangle =  0$.

We finish this part by showing how the analytic expressions recover those for the homogeneous spiked Wigner ensemble. Setting $P_S(s) = \delta(s - \sigma)$ in Eq. (\ref{q-final}), we obtain
the quadratic equation $\sigma q^2(z) - z q(z) + \sigma =0$, whose solutions read
\begin{equation}
  q_{\pm}(z) = \frac{1}{2 \sigma} \left( z \pm \sqrt{ z^2 - 4 \sigma^2  } \right).
  \label{ytrsd}
\end{equation}  
For $|\lambda| < 2 \sigma$, the root $q_{+}(z)$ leads to the Wigner semicircle law for $\rho(\lambda)$ with
support $\Omega_{\lambda} = (-2 \sigma,2 \sigma)$.

 
\subsection{Outside the bulk: spectral edge, outlier and the eigenvector distributions}

The calculation of the spectral edge $\lambda_*$ of $\rho(\lambda)$ is crucial for the inference problem under study. From Eq. (\ref{q-final}), we can derive an
equation for ${\rm Im} q(z)$,
\begin{equation}
  {\rm Im} q(z)\left(  \int_{\Omega_S} \frac{ds \, s^2 P_S(s)}{[\lambda- s \mkern1mu {\rm Re} q(z)]^2+[s \mkern1mu {\rm Im} q (z)]^2}-1  \right)=0.
  \label{tewds}
\end{equation}
For $\lambda$ outside the support $\Omega_{\lambda} = (-\lambda_{*}, \lambda_{*})$ of $\rho(\lambda)$, ${\rm Im} G_{ii}(z)$ converges to zero as $\epsilon \rightarrow 0^{+}$, implying
that $\lim_{\epsilon \rightarrow 0^{+}} {\rm Im}q(z)=0$. Thus, the above equation is trivially satisfied for $|\lambda| \geq |\lambda_{*}|$. On the other
hand, $\lim_{\epsilon \rightarrow 0^{+}} {\rm Im}q(z)$ is nonzero for $\lambda \in \Omega_{\lambda}$. As $|\lambda|$
approaches $|\lambda_{*}|$ from below, $\lim_{\epsilon \rightarrow 0^{+}} {\rm Im}q(z)$ becomes very small, and Eq. (\ref{tewds}) is satisfied provided 
\begin{equation}\label{edge}
    \int_{\Omega_S} \frac{ds \,s^2 P_S(s)}{\left[\pm \lambda_* - s \mkern1mu  q(\pm\lambda_*)    \right]^2} = 1.
\end{equation}
By solving Eq. (\ref{q-final}) and computing $ \lim_{\epsilon \to 0^+} {\rm Re} \mkern1mu q (z)$ as a function of $\lambda$, the above condition
determines the spectral edges $\pm \lambda_*$. Note that the spectral density and its support are independent of the
statistical properties of the spike $\boldsymbol{X}_n$, which means that a rank-one perturbation of $\mathcal{O}(n^{-1})$ does not affect the bulk spectrum of $\boldsymbol{A}_n$.

Let us now determine the outlier eigenvalue $\lambda_{\rm out}$ and the distribution $P_R(r|\lambda)$ of the eigenvector components for $\lambda \notin \Omega_{\lambda}$.
Building on the cavity approach as discussed in \cite{MNR2019}, we derive in appendix \ref{appecav} the
following distributional equation for an eigenvector component $R(\lambda)$ in the limit $n \rightarrow \infty$,
\begin{equation}\label{evecs}
  R(\lambda) \stackrel{\rm d}{=} \left[ \sqrt{\left\langle S
      R^2(\lambda) \right\rangle} \sqrt{S} U + \gamma \left\langle X R (\lambda) \right\rangle X \right] G(\lambda),
\end{equation}
where $U$ is a Gaussian
random variable with mean zero and variance one, while $G(\lambda)$ is computed from Eq. (\ref{juhrt}).
This equation is strictly valid outside the bulk.
The solutions of Eq. (\ref{evecs}) allow
us to study the statistics of the eigenvector components in terms of the model parameters, forming
an interesting starting point to compute the moments of $R(\lambda)$ and the outlier eigenvalue \cite{NeriMetz2016,MNR2019, Metz2021}.
From Eq. (\ref{evecs}), the first and second moments of $R(\lambda)$ fulfill the equations
\begin{eqnarray}\label{moms-evecs}
   \langle R (\lambda) \rangle & = & \gamma \langle X  R(\lambda)  \rangle \langle X\rangle \langle  G(\lambda)  \rangle, \\
  \label{moms-evecs2}
   \langle R^{2} (\lambda) \rangle & = &
  \langle S  R^{2}(\lambda) \rangle  \langle S G^{2}(\lambda) \rangle + \gamma^{2} \langle X R(\lambda) \rangle^{2} \langle X^2\rangle\langle G^{2}(\lambda) \rangle,
\end{eqnarray}
which depend on the correlations $\langle X R(\lambda) \rangle$ and $\langle S R^{2}(\lambda) \rangle$.
Thus, we derive from Eq. (\ref{evecs}) the additional expressions
\begin{eqnarray}
  \langle X R(\lambda) \rangle &=& \gamma  \langle X^2 \rangle \langle  G(\lambda) \rangle \langle X  R(\lambda)  \rangle,   \label{uycdd}  \\
  \langle X R(\lambda) \rangle^2 &=& \frac{\left[ 1-\langle S^2 G^{2}(\lambda)  \rangle \right]}{\gamma^2 \langle X^2 \rangle \langle S G^{2}(\lambda)  \rangle} \langle S R^{2}(\lambda) \rangle. \label{params}
\end{eqnarray}  
The solutions of the fixed-point Eqs. (\ref{moms-evecs}-\ref{params}) yield the outlier $\lambda_{\rm out}$, the spectral edge $\lambda_{*}$, and
the first two moments of the corresponding eigenvector components.

First, let us consider the trivial solution $\langle X R(\lambda)\rangle = 0$ of Eq. (\ref{uycdd}), which represents the absence of correlation between
the eigenvector and the spike. By setting  $\langle X R(\lambda) \rangle = 0$ in Eq. (\ref{params}), we find that the right hand side of this equation is
zero for values of $\lambda$ such that $\langle S^2 G^{2}(\lambda)\rangle = 1$. This condition coincides with Eq. (\ref{edge}), which determines the spectral edges. Hence, the solution
$\langle X R(\lambda) \rangle = 0$ appears strictly at $\lambda = \pm \lambda_*$, and the first two moments of the eigenvector
components at the spectral edges are given by $\langle R(\pm \lambda_*) \rangle =0$
and $\langle R^{2}(\pm \lambda_*) \rangle =  \langle S R^{2}(\pm \lambda_*) \rangle\langle S G^2(\pm \lambda_*) \rangle$.

Next, we consider the case in which the eigenvector components have a nonzero correlation $\langle X R(\lambda) \rangle$ with the spike. The
linear Eq. (\ref{uycdd}) admits a nontrivial solution, $\langle X R(\lambda) \rangle \neq 0$, when $\lambda=\lambda_{\rm out} > 0$ satisfies
\begin{equation}\label{outlier}
     \int_{\Omega_S} \frac{ds P_S(s)}{\left[ \lambda_{\rm out} - s  q(\lambda_{\rm out})     \right]} = \frac{1}{ \gamma\langle X^2\rangle }.
\end{equation}
If the solution of this equation detaches from the bulk ($\lambda_{\rm out} > \lambda_{*}$), the spectrum of $\boldsymbol{A}_n$ exhibits an outlier at $\lambda=\lambda_{\rm out}$. In this
regime, the signal $\boldsymbol{X}_n$ is detectable, since the spectrum of $\boldsymbol{A}_n$ can be distinguished from the spectrum of the
noise matrix in Eq. (\ref{model}). The transition between
the gapped phase, where the spectral gap $\lambda_{\rm out} - \lambda_{*}$ is finite, and the gapless phase, where the spectrum of $\boldsymbol{A}_n$
has no outlier, is known as the BBP transition \cite{BBP}. Here, we can compute the BBP transition line by identifying the set of parameters
for which $\lambda_{\rm out} = \lambda_{*}$ solves simultaneously Eqs. (\ref{edge}) and (\ref{outlier}). From Eqs. (\ref{moms-evecs}) and (\ref{moms-evecs2}), we
derive the first two moments of the eigenvector components associated to the outlier
\begin{eqnarray}
  \frac{\langle R (\lambda_{\rm out}) \rangle}{\langle X  R(\lambda_{\rm out})  \rangle } & = & \gamma  \langle X\rangle \langle  G(\lambda_{\rm out})  \rangle, \label{mom1} \\
  \frac{\langle R^{2}(\lambda_{\rm out})\rangle  }{ \langle X R(\lambda_{\rm out}) \rangle^2  }  &=&
  \frac{\gamma^2 \langle X^2\rangle \langle S  \mkern1mu G^{2}(\lambda_{\rm out})  \rangle}{\left[ 1-\langle S^2 G^{2}(\lambda_{\rm out})  \rangle \right]} \langle S  \mkern1mu G^{2}(\lambda_{\rm out}) \rangle
    + \gamma^2 \langle X^2\rangle \langle G^{2} (\lambda_{\rm out})\rangle . \label{mom2}
\end{eqnarray}
Since the eigenvectors are defined up to an overall normalization, one can fix either $\langle R (\lambda_{\rm out}) \rangle$ or $\langle R^{2}(\lambda_{\rm out})\rangle$, which
then fully determines the first and second moments through Eqs. (\ref{mom1}) and (\ref{mom2}).

The components of the outlier eigenvector provide a natural estimator for $\boldsymbol{X}_n$ in the gapped phase.
From the distributional Eq. (\ref{evecs}), one can derive an explicit expression for the distribution
of the eigenvector components corresponding to eigenvalues outside the bulk,
\begin{equation}
  P_R(r | \lambda) = \int_{\mathbb{R}} d x P_X(x) \int_{\Omega_{S}} d s P_S(s) \frac{1}{\sqrt{2 \pi \sigma_{R}^{2}(s|\lambda)  }} \exp{\left( - \frac{1}{2 \sigma_{R}^{2}(s|\lambda) }
      \left[ r - \mu(s,x | \lambda)  \right]^2 \right) },
  \label{ured}
\end{equation}  
where
\begin{equation}
\mu(s,x|\lambda) = \frac{\gamma \langle X R(\lambda)  \rangle x }{\lambda - s \langle S G(\lambda)  \rangle }
\end{equation}  
and
\begin{equation}
\sigma_{R}^{2}(s|\lambda) = \frac{\langle S R^{2}(\lambda) \rangle s  }{\left[ \lambda - s \langle S G(\lambda)  \rangle \right]^2},
\end{equation}  
with $\lambda = \pm \lambda_{*}$ or $\lambda=\lambda_{\rm out}$. For $\lambda \notin \Omega_{\lambda}$, the distribution of the eigenvector
components is thus a weighted superposition of Gaussian distributions with mean and variance
governed by the spike prior $P_X$ and the variance profile $P_S$.
Note that $P_R(r|\lambda)$ is fully specified once the correlation $\langle X R(\lambda)  \rangle$ is fixed, reflecting the freedom in the normalization of the eigenvectors. 
Equation (\ref{ured}) provides a natural prescription for estimating the rank-one signal
in the gapped phase ($\lambda = \lambda_{\rm out}$). We can access the quality of this estimate by computing the mean squared error
\begin{equation}\label{mse}
  {\rm MSE}(\boldsymbol{X}|\boldsymbol{R}) = \lim_{n \rightarrow \infty}  \frac{1}{n} \sum_{i=1}^n \left[ X_{i,n} - R_{i,n}(\lambda_{\rm out}) \right]^{2}
  = \langle X^2  \rangle + \langle R^{2}(\lambda_{\rm out}) \rangle - 2 \langle X R(\lambda_{\rm out})  \rangle.
\end{equation}  
Since $\langle X R(\pm \lambda_*)  \rangle = 0$ and $\mu(s,x|\pm \lambda_*) = 0$, $P_R(r | \pm \lambda_*)$ is independent
of the spike distribution $P_X$ at the spectral edges. Consequently, the eigenvectors at $\lambda = \pm \lambda_*$ contain
no information about the spike $\boldsymbol{X}_n$.

We now recover well-known results for the homogeneous spiked Wigner model from our general equations. Setting $P_S(s) = \delta( s - \sigma)$
and choosing the root ${\rm Re} q_{+}(\lambda) = \lambda/2 \sigma$ in Eq. (\ref{ytrsd}), the solution of Eq. (\ref{edge}) gives
the spectral edge $\lambda_{*}=2 \sigma$ of the Wigner semicircle law. To obtain the outlier from Eq. (\ref{outlier}), we instead
select the root ${\rm Re} q_{-}(\lambda) = \frac{1}{2 \sigma} (\lambda - \sqrt{\lambda^2 - 4 \sigma^2} )$ in Eq. (\ref{ytrsd}), which
satisfies $\lim_{\lambda \rightarrow \infty} {\rm Re} q_{-}(\lambda) = 0$ and therefore provides the physically relevant
solution for $q(\lambda)$ outside the bulk. Substituting $P_S(s) = \delta( s - \sigma)$
and ${\rm Re} q_{-}(\lambda)$ into Eq. (\ref{outlier}), we recover the outlier eigenvalue of the
homogeneous spiked Wigner model, Eq. (\ref{out-wigner}). From the condition $\lambda_* = \lambda_{\rm out}$, we find the quadratic
equation $\gamma^2 \langle X^2 \rangle^2 - 2 \sigma \gamma \langle X^2 \rangle + \sigma^2 = 0$ for $\gamma \langle X^2 \rangle$, whose solution
yields the BBP transition line $\gamma \langle X^2 \rangle = \sigma$, Eq. (\ref{snr-wigner}).
The distribution  of the eigenvector components
of the homogeneous Wigner model follows by setting  $P_S(s) = \delta(s - \sigma)$  in Eq. (\ref{ured}).
For $\lambda = \pm \lambda_*$, the eigenvector components follow the Gaussian distribution
\begin{equation}
  P_R(r | \pm \lambda_*) =  \frac{1}{\sqrt{2 \pi \langle R^{2}(\pm \lambda_*) \rangle   }} \exp{\left[ - \frac{r^2}{2  \langle R^{2}(\pm \lambda_*) \rangle  }   \right]},
  \label{ured1}
\end{equation}  
with $\lambda_{*}=2 \sigma$. For $\lambda = \lambda_{\rm out}$ (see Eq. (\ref{out-wigner})), we obtain
\begin{equation}
  P_R(r | \lambda_{\rm out}) = \int_{\mathbb{R}} d x P_X(x) \frac{1}{\sqrt{2 \pi \sigma_{R}^{2}(\lambda_{\rm out})  }}
  \exp{\left( - \frac{1}{2 \sigma_{R}^{2}(\lambda_{\rm out}) }  \left[ r - \mu(x|\lambda_{\rm out})  \right]^2 \right) },
  \label{ured2}
\end{equation}  
where
\begin{equation}
  \mu(x|\lambda_{\rm out}) = \frac{\langle X R(\lambda_{\rm out})  \rangle x }{\langle X^2\rangle} \quad {\rm and}
\quad \sigma_{R}^{2}(\lambda_{\rm out}) = \frac{\sigma^2 \langle R^{2}(\lambda_{\rm out}) \rangle  }{(\gamma\langle X^2\rangle)^2}.
\end{equation}  

Finally, we determine the empirical distribution $P_G(g|\lambda)$ of the resolvent entries $G_{ii}(\lambda)$ for $|\lambda| \geq |\lambda_*|$. Although this object is not directly
relevant for the inference problem considered here, $P_G(g|\lambda)$ plays an important role in random-matrix theory and the study of localization
phenomena \cite{Mirlin2000,Jeferson2022,Tapias2023,Silva2025}, as it determines the local density of states.
For $|\lambda| \geq |\lambda_*|$, the distribution $P_G(g|\lambda)$ is
supported on the real line $g \in \mathbb{R}$. From the nonlinear relation between $G_{ii}$ and $S_i$, given by Eq. (\ref{juhrt}), one can perform a change of
variables and obtain, for arbitrary $P_S$ with support $\Omega_{S} \in [s_{-},s_{+}]$, the following expression
\begin{equation}\label{out-res-dist}
    P_{G}(g|\lambda) = \frac{1}{|q(\lambda)| g^2} P_S\left[ \frac{\lambda}{q(\lambda)} - \frac{1}{q(\lambda)g}\right].
\end{equation}
For $\lambda > 0$ outside the bulk, the support of $P_{G}$ reads
$\Omega_G \in [\frac{1}{\lambda - q(\lambda) s_{-} }, \frac{1}{\lambda - q(\lambda) s_{+} } ]$, with $q(\lambda) > 0$. For $\lambda < 0$ outside
the bulk, the support of $P_{G}$ is given by
$\Omega_G \in [\frac{1}{\lambda - q(\lambda) s_{+} }, \frac{1}{\lambda - q(\lambda) s_{-} } ]$, with $q(\lambda) < 0$.


\section{Numerical results}
\label{power-law}

In the previous section, we have shown how to determine the spectral properties of the observation matrix $\boldsymbol{A}_n$ in
the limit $n\to\infty$. The equations presented in section \ref{derivation} are valid for an arbitrary spike distribution $P_X$ and variance
profile $P_S$ with finite support $\Omega_S$.
In this section, we examine the effects of noise inhomogeneity on the detection and recovery of the spike for specific
choices of $P_X$ and $P_S$. In order to reduce numerical instabilities, we evaluate the integrals
  in Eqs. (\ref{edge}) and (\ref{outlier}) for $z=\lambda-i\epsilon$ and
  compute $q(\lambda)$ outside the bulk from the limit $\lim_{\epsilon\to 0^+} {\rm Re} \,q(\lambda-i\epsilon)$.

For the variance profile, we consider the following truncated power-law distribution
\begin{equation} \label{ps}
P_S(s) =
\left\{
\begin{array}{ll}
  \mathcal{A}^{-1} s^{\alpha}, & {\rm if} \,\, s \in \left[1-\frac{\Delta}{2},1+\frac{\Delta}{2}\right], \\
0,   & {\rm otherwise},
\end{array}
\right.
\end{equation}
where $\mathcal{A}$ is the normalization constant 
\begin{equation}
    \mathcal{A} = \frac{1}{\alpha+1} \left[\left(1+\frac{\Delta}{2}\right)^{\alpha+1}-\left(1-\frac{\Delta}{2}\right)^{\alpha+1}\right].
\end{equation}
The parameter $0 < \Delta \leq 2$ controls the width of the support $\Omega_S=\left[1-\frac{\Delta}{2},1+\frac{\Delta}{2}\right]$, while
the exponent $-1 < \alpha < \infty$ governs the shape of $P_S$. The mode of $P_S(s)$ is located at $s=1+\Delta/2$ or $s=1-\Delta/2$ provided $\alpha > 0$ or $\alpha < 0$, respectively. 
For $\alpha=0$, Eq. (\ref{ps}) reduces to the uniform distribution.
The first and second moments of $P_S$ are given by
\begin{equation}\label{moms-s}
  \langle S^{m} \rangle = \frac{\alpha+1}{\alpha+m+1}
  \frac{\left[\left(1+\frac{\Delta}{2}\right)^{\alpha+m+1}-\left(1-\frac{\Delta}{2}\right)^{\alpha+m+1} \right]}{\left[\left(1+\frac{\Delta}{2}\right)^{\alpha+1}-\left(1-\frac{\Delta}{2}\right)^{\alpha+1} \right]}, 
\end{equation}
with $m=1,2$. The above expression is useful to understand the behaviour of $P_S$ in certain limiting cases:
\begin{itemize}
\item As $\alpha\to\infty$, both $\langle S \rangle^2,\langle S^2\rangle \to \left(1+\frac{\Delta}{2}\right)^2$, thus
  $P_S(s)\to\delta[s-(1+\frac{\Delta}{2})]$, and we recover the homogeneous spiked Wigner ensemble with standard deviation $\sigma=1+\frac{\Delta}{2}$.

\item As $\alpha\to-1$ and $\Delta \to 2$,  both $\langle S \rangle^2,\langle S^2\rangle \to 0$, and hence $P_S(s)\to\delta(s)$. In this case, the noise component
  in Eq. (\ref{model}) is zero.
\end{itemize}
For $\Delta =0$, we recover the homogeneous spiked Wigner ensemble with $\sigma=1$, since $P_S(s) = \delta(s-1)$.
This variance profile thus enables to continuously interpolate
  between different limiting regimes and
to investigate the effect of both the
      typical magnitude of the variables $\{ S_i \}_{i=1}^N$ and the strength of their fluctuations on the spectral
      properties. Regarding the
spike distribution $P_X$, below we present results for a constant spike, where $P_X(x) = \delta(x-1)$, and for
spike components drawn from a Gaussian distribution.


\subsection{Empirical spectral density}

\begin{figure}[h!]
    \centering
    \includegraphics[scale=0.8]{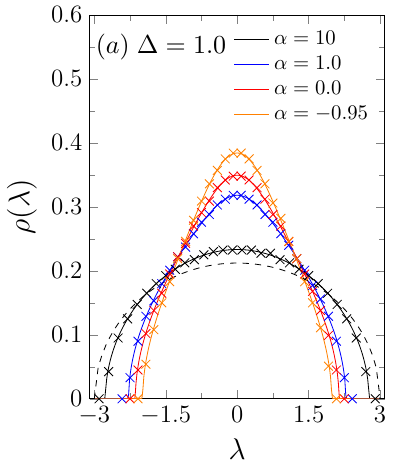}
    \includegraphics[scale=0.8]{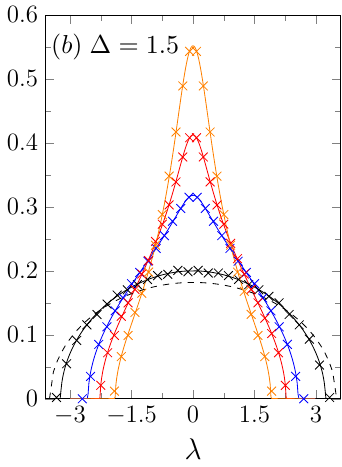}
    \includegraphics[scale=0.8]{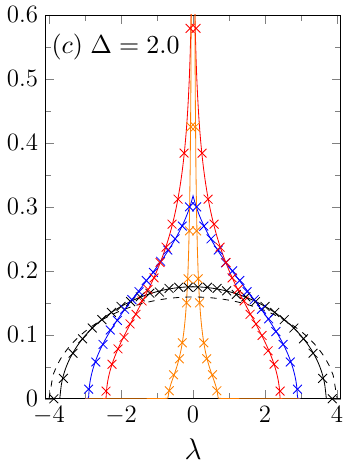}
    \caption{Empirical spectral density $\rho(\lambda)$ of the observation matrix $\boldsymbol{A}_n$ (see Eq. (\ref{model})) with a truncated
      power-law variance profile $P_S(s)$ defined in Eq. (\ref{ps}). Solid lines are the solutions of Eqs. (\ref{q-final}) and (\ref{ESD-final})
      with $\epsilon=10^{-3}$, for different values of the width parameter $\Delta$ and exponent $\alpha$, as indicated in the figures. Dashed
      lines represent the Wigner semicircle law with $\sigma=1+\frac{\Delta}{2}$ (see Eq. (\ref{wigner})). Symbols are results obtained
      from the numerical diagonalization of an ensemble of 100 matrices $\boldsymbol{A}_n$ generated according to Eq. (\ref{model}), with signal-to-noise
      ratio $\gamma=0$ and dimension $n=2000$.}
    \label{fig:2}
\end{figure}

The spectral density $\rho(\lambda)$ of $\boldsymbol{A}_n$ is obtained from the solutions of Eqs. (\ref{q-final}) and (\ref{ESD-final}).
Figure \ref{fig:2} illustrates the behaviour of $\rho(\lambda)$ for the truncated power-law distribution $P_S(s)$ defined
in Eq. (\ref{ps}). As $\alpha \to \infty$, $P_{S}(s)$ develops a peak at the upper edge of its support, and $\rho(\lambda)$ converges to the Wigner semicircle
law $\rho_{\rm sc}(\lambda)$ with $\sigma=1+\frac{\Delta}{2}$ (see Eq. (\ref{wigner})). In contrast, as $\alpha\to-1$, the weight $P_{S}(s)$
near the lower edge of the support $\Omega_S$ increases, leading to an overall reduction in the variance of the  noise contribution in Eq. (\ref{model}).
Accordingly, the support of $\rho(\lambda)$ shrinks as $\alpha\to-1$, vanishing in the particular case of $\Delta=2$ (figure \ref{fig:2}$(c)$).
Figure \ref{fig:2} also shows that results from numerical diagonalizations of finite-dimensional matrices $\boldsymbol{A}_n$ are in excellent agreement with
the solutions of Eqs. (\ref{q-final}) and (\ref{ESD-final}), thus confirming our theoretical predictions for $n \rightarrow \infty$.

Figure \ref{fig:2}$(c)$ shows that $\rho(\lambda)$ may exhibit a divergence  at $\lambda=0$ when $\Delta=2$. This singular behaviour originates
from strong fluctuations in the imaginary parts of the
diagonal elements of the resolvent \cite{Jeferson2022,Tapias2023,Silva2025}. Defining $Y_{n,i}= {\rm Im} G_{n,ii}(0) \geq 0$, one can
analytically determine, in the limit  $n \rightarrow \infty$, the empirical distribution $P_Y(y)$ of $\{ Y_{n,i} \}_{i=1}^n$. For $P_S(s)$ given by Eq. (\ref{ps}), one
finds that ${\rm Im} q(0) = 1$, and the distribution $P_Y(y)$ reads
\begin{equation} \label{ldos}
P_Y(y) =
\left\{
\begin{array}{ll}
\mathcal{A}^{-1} y^{-(\alpha+2)} , & {\rm if} \,\,  y \in \Omega_Y, \\
0,   & {\rm otherwise},
\end{array}
\right.
\end{equation}
where $\Omega_Y = \left[\left(1+\frac{\Delta}{2}\right)^{-1},\left(1-\frac{\Delta}{2}\right)^{-1} \right]$ denotes the support of $P_Y(y)$.
According to Eq. (\ref{ESD-sum}), the spectral density $\rho(\lambda)$ at $\lambda=0$ is proportional to the first
moment $\langle Y \rangle = \int_{\Omega_{Y}} dy \,  y \, P_Y(y)$. For $\Delta=2$, this gives
\begin{equation}
    \langle Y \rangle = \frac{1}{\mathcal{A}}\int_{1/2}^{\infty} dy\, y^{-(\alpha+1)}.
\end{equation}
The above expression diverges for $\alpha \leq 0$, in full agreement with the results in figure \ref{fig:2}$(c)$.
This singularity therefore stems from strong fluctuations in the diagonal elements of the resolvent, which are induced
by arbitrarily small variances of the individual entries in the noise component of Eq. (\ref{model}). As shown in previous studies \cite{Jeferson2022,Tapias2023}, this
heterogeneity-induced divergence within the bulk of the spectrum is associated  with localization of the corresponding eigenvectors on sites with small values of $S_i$.


\subsection{Spectral edge and outlier eigenvalue}

\begin{figure}[h!]
    \centering
    \includegraphics[scale=0.85]{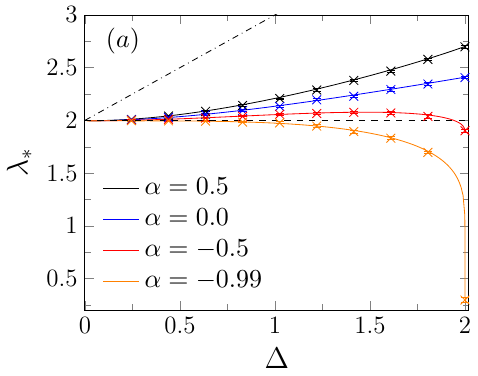}
    \includegraphics[scale=0.85]{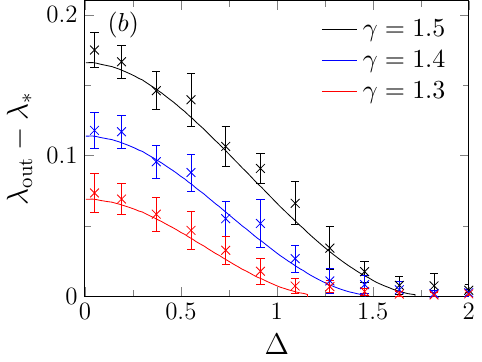}
    \caption{Spectral edge and spectral gap of the observation matrix $\boldsymbol{A}_n$, obtained from
      the solutions of Eqs. (\ref{q-final}) and (\ref{edge}) (solid lines) for small values of $\epsilon$ and a deterministic spike ($P_X(x)=\delta(x-1)$).
      The variance profile $P_S(s)$ is given by a truncated power-law distribution with width $\Delta$ and shape parameter $\alpha$ (see Eq. (\ref{ps})).
      The symbols represent average results obtained from the numerical diagonalization of an ensemble of $10$ matrices
      generated according to Eq. (\ref{model}) with dimension $n=5000$, while the vertical bars are the corresponding standard deviations.
      (a) Spectral edge $\lambda_*$ as a function of $\Delta$ for $\gamma=0$.
      The dash-dotted line shows the edge $\lambda_{*} = 2 (1 + \Delta/2)$ of the homogeneous spiked Wigner model with $\sigma = 1 + \Delta/2$ (see Eq. (\ref{wigner})).
      (b) Spectral gap $\lambda_{\rm out}-\lambda_*$ as a function of $\Delta$ for $\alpha=1/2$.}
    \label{fig:3}
\end{figure}

The spectral edge $\lambda_{*} > 0$ follows from the solutions of Eqs. (\ref{q-final}) and (\ref{edge}).
In figure \ref{fig:3}$(a)$, we present results for different combinations of the parameters $\alpha$ and $\Delta$
that characterize the truncated power-law distribution $P_S$ (see Eq. (\ref{ps})).
As $\Delta \to 0$, fluctuations in the variance matrix $\boldsymbol{S}_n$ are suppressed and $\lambda$ converges
to the spectral edge $\lambda_{*} = 2$ of the homogeneous spiked Wigner model.
For $\alpha \geq 0$, $\lambda_{*}$ increases monotonically with $\Delta$.
In the limit $\alpha \to \infty$, we recover the spectral edge $\lambda_{*} = 2 (1 + \Delta/2)$ of the homogeneous
spiked Wigner model with $\sigma = 1 + \Delta/2$, since $P_S(s) = \delta[ s - (1 + \Delta/2   ) ]$.
For moderate negative values of $\alpha$, when the mode of $P_S(s)$ is located at the lower
edge $1 - \Delta/2$, $\lambda_{*}$ becomes a non-monotonic function of $\Delta$.
Setting $\Delta = 2$ and taking the limit $\alpha \to -1$ yields $P_S(s) = \delta(s)$, which
results in $\lambda_{*} \to 0$.

The spectral gap $\lambda_{\rm out}-\lambda_{*}$ follows from the solutions of Eqs. (\ref{edge}) and (\ref{outlier}).
In figure \ref{fig:3}$(b)$, we compare the solutions of these equations with numerical results from
the diagonalization of finite-dimensional matrices $\boldsymbol{A}_n$ for a deterministic spike
prior $P_X(x)=\delta(x-1)$, confirming our theoretical predictions in the limit $n \to \infty$.
When $\lambda_{\rm out} - \lambda_{*} > 0$, the spectrum of $\boldsymbol{A}_n$ is gapped and the signal
$\boldsymbol{X}_n$ is detectable through the outlier eigenpair.
When $\lambda_{\rm out} - \lambda_{*} \leq 0$, the spectrum is gapless and the spike vector cannot be
detected by spectral methods. The vanishing of the spectral gap identifies the BBP transition line \cite{BBP}.
Figure \ref{fig:3}$(b)$ shows that, for $\alpha > 0$ and moderate values of the
signal strength $\gamma$, the spectral gap vanishes monotonically as the width $\Delta$ of the support $\Omega_S$ increases.

\subsection{Eigenvectors outside the bulk}

\begin{figure}[h!]
    \centering
    \includegraphics[scale=0.85]{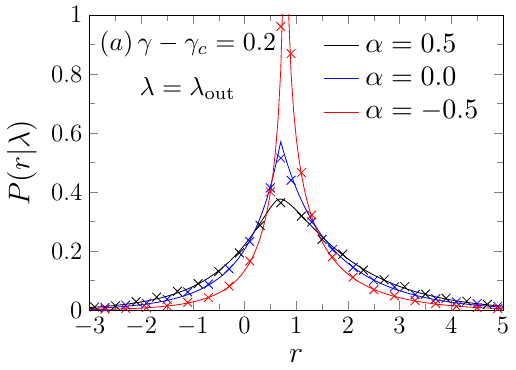}
    \includegraphics[scale=0.85]{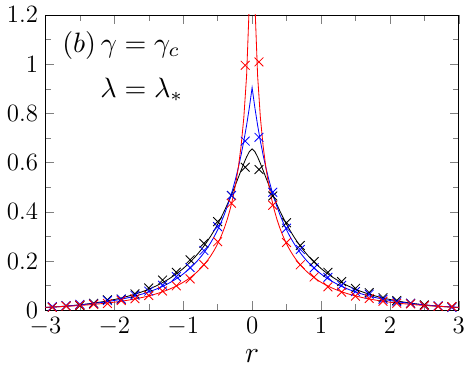}
    \caption{Distribution of the eigenvector components outside the bulk of eigenvalues, obtained from Eq. (\ref{ured}) (solid lines)
      for a deterministic spike ($P_X(x)=\delta(x-1)$). The variance profile $P_S(s)$ is given by a truncated power-law
      distribution with width $\Delta=2$ and shape parameter $\alpha$ (see Eq. (\ref{ps})). The BBP threshold $\gamma_c=\gamma_c(\Delta,\alpha)$ is
      computed from the solutions of Eqs. (\ref{edge}) and (\ref{outlier}). Symbols are results obtained from the numerical diagonalization of an ensemble of $10$ matrices
      generated according to Eq. (\ref{model}) with dimension $n=5000$. (a) Eigenvector distribution $P_R(r|\lambda_{\rm out})$ for $\langle R\rangle =1$, corresponding
      to the outlier eigenvalue. (b) Eigenvector distribution $P_R(r|\pm \lambda_{*})$ for $\langle R^2 \rangle =1$, corresponding
      to the spectral edges.}
    \label{fig:4}
\end{figure}

The distribution $P_R(r|\lambda)$ of the eigenvector components outside the bulk follows directly from Eq. (\ref{ured}).
In figure \ref{fig:4}, we illustrate the behavior of $P_R(r|\lambda)$ for a deterministic spike prior $P_X(x)=\delta(x-1)$
and a truncated power-law distribution $P_S(s)$ with maximum width $\Delta=2$ and different shape parameters $\alpha$.
For each pair $(\Delta,\alpha)$, the spectrum of $\boldsymbol{A}_n$ undergoes a BBP transition at a critical signal-to-noise ratio $\gamma=\gamma_c$.
When the signal is sufficiently strong ($\gamma > \gamma_c$), the spectrum contains an outlier $\lambda_{\rm out}$, and
figure \ref{fig:4}$(a)$ shows the corresponding eigenvector distribution $P_R(r|\lambda_{\rm out})$ for the choice $\langle R \rangle = \langle X \rangle$.
For $\gamma \leq \gamma_c$, no outlier is present, and Eq. (\ref{ured}) yields the eigenvector
distribution $P_R(r|\pm \lambda_{*})$ at the spectral edges, shown in figure \ref{fig:4}$(b)$ for the choice $\langle R^2\rangle=1$.
The eigenvector distributions are non-Gaussian, in contrast to the Gaussian behavior of the homogeneous spiked Wigner
ensemble (see Eqs. (\ref{ured1}) and (\ref{ured2})).
Figure \ref{fig:4} further shows that, in both phases (gapped and gapless), the distribution $P_R(r|\lambda)$ concentrates
around $\langle R\rangle$ as $\alpha$ decreases.
This occurs because the fluctuations of $R_i(\lambda)$ around the spike
component $X_i$ are controlled by $\sqrt{S_i}$, which become smaller as $\alpha \to -1$.
Finally, numerical results for the distribution of the eigenvector components, obtained from the diagonalization of $\boldsymbol{A}_n$ for
finite $n$, show excellent agreement with Eq. (\ref{ured}) in both regimes displayed in figure \ref{fig:4}.

\begin{figure}[h!]
    \centering
    \includegraphics[scale=0.75]{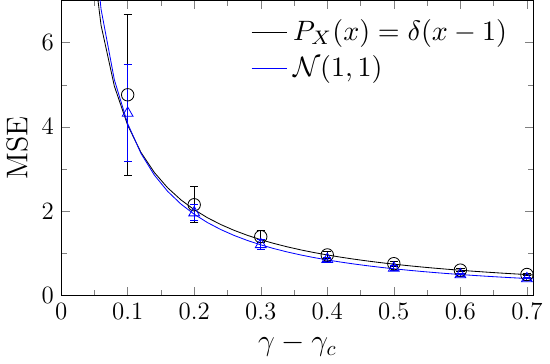}
    \caption{Mean squared error $\rm MSE$ between the spike vector $\boldsymbol{X}_n$ and the eigenvector
      $\boldsymbol{R}_n(\lambda_{\rm out})$ associated to the outlier eigenvalue of the observation matrix $\boldsymbol{A}_n$, for both deterministic and
      Gaussian-distributed spikes. The variance profile $P_S(s)$ follows a truncated power-law distribution
      with $\Delta=1$ and $\alpha=0.5$. The BBP threshold $\gamma_c=\gamma_c(\Delta,\alpha)$ is
      computed from the solutions of Eqs. (\ref{edge}) and (\ref{outlier}). Solid
      lines represent analytic results from Eq. (\ref{mse}). Symbols are average results obtained from the
      numerical diagonalization of an ensemble of $10$ matrices generated according to Eq. (\ref{model}) with $n=5000$, while
    the vertical bars are the corresponding standard deviations.}
    \label{fig:5}
\end{figure}

Figure \ref{fig:4}$(a)$ shows that the eigenvector components associated with $\lambda_{\rm out}$ are
moderately spread around the mean $\langle X \rangle = 1$ of the spike components.
Thus, the outlier eigenvector $\boldsymbol{R}_n(\lambda_{\rm out})$ provides
an estimate of the rank-one signal $\boldsymbol{X}_n$ in the gapped regime.
The quality of this estimate can be measured by the mean squared error ${\rm MSE}(\boldsymbol{X}|\boldsymbol{R})$
between $\boldsymbol{X}_n$ and $\boldsymbol{R}_n(\lambda_{\rm out})$, given by Eq. (\ref{mse}) in the limit $n \to \infty$.
Figure \ref{fig:5} shows the behavior of the ${\rm MSE}$ as a function of $\gamma$ for both deterministic and
random spikes, with $X_{n,i} = 1$ and $X_{n,i} \sim \mathcal{N}(1,1)$, respectively. For both spike distributions, we
  choose the normalization $\langle R(\lambda) \rangle = 1$.
The ${\rm MSE}$ vanishes as $\gamma \to \infty$, whereas it diverges as the BBP 
threshold $\gamma = \gamma_c$ is approached. This divergence confirms that the eigenvector at
the spectral edge does not carry information about the spike.

\subsection{The BBP transition}

\begin{figure}[H]
    \centering
    \includegraphics[scale=0.7]{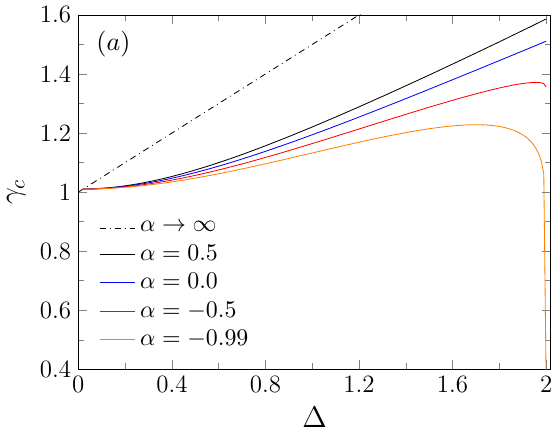}
    \includegraphics[scale=0.7]{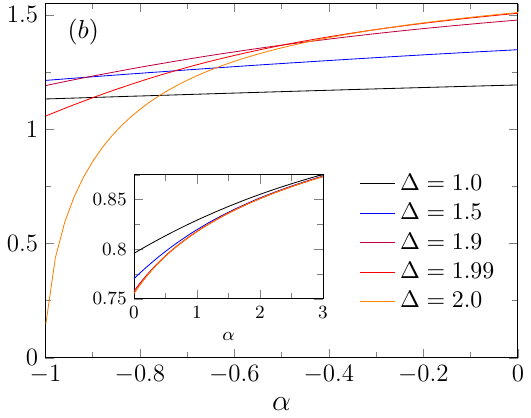}  
    \caption{Critical value $\gamma_c$ of the signal-to-noise ratio marking the BBP transition (see the main text). When $\gamma > \gamma_c$, the spectral gap
      of the observation matrix is finite and the rank-one signal can be detected. These results are
      obtained from the solutions of Eqs. (\ref{edge}) and (\ref{outlier}) for $\epsilon=10^{-3}$
      and a truncated power-law distribution $P_S(s)$  with width $\Delta$ and shape parameter $\alpha$ (see Eq. (\ref{ps})). The spike is
      generated from the distribution $P_X(x) = \delta(x-1)$. The dash-dotted straight line $\gamma_c = 1 + \Delta/2$ (left panel) is
      the limiting behaviour for $\alpha \rightarrow \infty$. The inset (right panel) shows the rescaled quantity $\gamma_c/(1 + \Delta/2)$ as a function
      of $\alpha \geq 0$. }
    \label{fig:6}
\end{figure}

The critical signal-to-noise ratio $\gamma_c$ that identifies
the BBP transition is obtained by solving simultaneously Eqs. (\ref{edge}) and (\ref{outlier}).
Figures \ref{fig:6}$(a)$ and \ref{fig:6}$(b)$ show $\gamma_c(\alpha,\Delta)$ as a function of, respectively,  $\Delta$ and $\alpha$, the parameters
that characterize the variance profile $P_S(s)$ defined in Eq. (\ref{ps}).
For $\gamma > \gamma_c(\alpha,\Delta)$, the spectrum of the observation matrix $\boldsymbol{A}_n$ contains
an outlier separated from the bulk by a finite gap, and the signal can be estimated from the associated eigenvector.
For $\gamma \leq \gamma_c(\alpha,\Delta)$, the spectrum of $\boldsymbol{A}_n$ is gapless and
the signal is not detectable by spectral methods.
As shown in figure \ref{fig:6}, introducing randomness in the individual variances of the noise matrix makes signal detection more
difficult for $\alpha \geq  0$, since $\gamma_c$ grows monotonically with $\Delta$.
In contrast, for $\alpha < 0$, when the mode of $P_S(s)$ lies at the lower edge $1-\Delta/2$, $\gamma_c$ displays a pronounced non-monotonic dependence on $\Delta$.
In particular, when $\alpha \simeq -1$,  $\gamma_c$ decreases rapidly as $\Delta \rightarrow 2$, as a consequence of the
  arbitrarily small variances in the noise matrix. This non-monotonic behavior shows that fluctuations
in the variances of the noise matrix can also enhance signal detection.
Figure \ref{fig:6} also shows that $\gamma_c \rightarrow 1 + \Delta/2$ in the limit $\alpha \rightarrow \infty$, recovering
the BBP threshold of the homogeneous spiked Wigner model with $\sigma = 1 + \Delta/2$ (see Eq. (\ref{snr-wigner})).

\begin{figure}[H]
    \centering
    \includegraphics[scale=0.7]{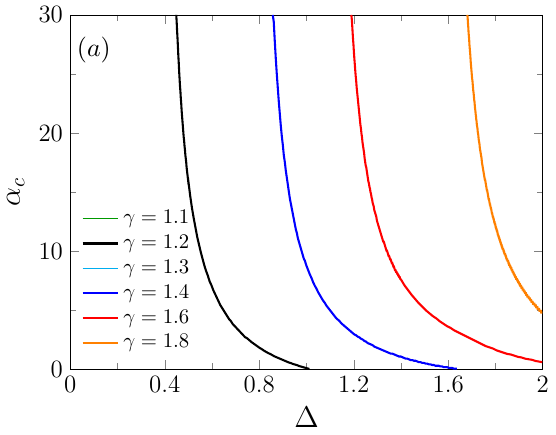}
    \includegraphics[scale=0.7]{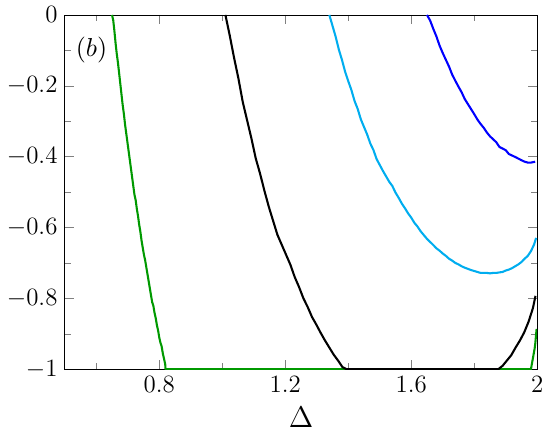}
    \caption{Critical value $\alpha_c$ of the shape parameter marking the BBP transition (see the main text).
      These results are  obtained from the solutions of Eqs. (\ref{edge}) and (\ref{outlier}) for $\epsilon=10^{-3}$
      and a truncated power-law distribution $P_S(s)$  with width $\Delta$ and shape parameter $\alpha$ (see Eq. (\ref{ps})).
      When $\alpha < \alpha_c$, the spectral gap
      of the observation matrix is finite and the rank-one signal can be detected.  The spike is
      generated from the distribution $P_X(x) = \delta(x-1)$.}
    \label{fig:7}
\end{figure}

Lastly, we discuss the BBP critical line in the $(\Delta,\alpha)$ plane.
In figure \ref{fig:7}, we present results for the critical value $\alpha_c(\Delta,\gamma)$
below which the spectrum of $\boldsymbol{A}_n$ is gapped and the signal is detectable.
Figure \ref{fig:7}$(b)$ shows that $\alpha_c(\Delta,\gamma)$ displays a non-monotonic dependence
on $\Delta$ for negative $\alpha$. In particular, for low values of the signal strength $\gamma$, a moderate
increase of the width parameter $\Delta$ completely suppresses the gapped phase. However, upon further increasing $\Delta$, the gapped
phase re-emerges near $\Delta = 2$, due to the overall reduction of the noise caused by the
many near-zero variances of the noise matrix entries in Eq. (\ref{model}).
For positive $\alpha$, where the mode of $P_S$ lies at the right edge of its support, $\alpha_c(\Delta,\gamma)$
decreases monotonically with $\Delta$.

\section{Summary and conclusions} \label{conclusion}

The spiked Wigner ensemble \cite{Peche2006} is one of the simplest models for studying the inference of a
low-rank signal corrupted by additive noise. In its standard homogeneous form, the entries of the noise matrix
are independent and identically distributed random variables. In this work, we have investigated how inhomogeneities in the noise matrix shape
the spectral properties of the spiked Wigner model.
To this end, we have considered an analytically tractable extension in which the noise component is given by the Hadamard
product of a Wigner matrix and a variance matrix with factorized entries of the form $S_i S_j$, where the variables $\{ S_i \}_{i=1}^n$ are
drawn from an arbitrary distribution that defines the variance profile. This construction provides a simple way to introduce nontrivial noise inhomogeneity. It
also arises naturally as the high-connectivity limit of sparse random matrices encoding the structure of random
graphs \cite{Metz2020,Jeferson2022}.
In the high-dimensional limit, we have derived exact equations for the spectral properties of the observation matrix relevant
to inference: the empirical spectral density and its support, the outlier eigenvalue, the BBP transition
line, and the full distribution of the eigenvector components associated with the outlier.

When the spectrum of the observation matrix contains an outlier eigenvalue detached from the bulk, the corresponding eigenvector can be used to estimate the rank-one
signal. One of our main results is an analytic expression
for the full distribution of the outlier eigenvector components. This distribution is a weighted superposition of Gaussian
distributions, with mean and variance determined by both the signal prior and the variance profile. The eigenvector distribution 
thus provides a direct way to estimate the signal in the gapped regime, without requiring more sophisticated algorithms.
At the BBP transition, the outlier merges with the spectral bulk and the estimator
based on the leading eigenvector fails, since the eigenvectors at the spectral edges
carry no information about the spike. We showed this explicitly by deriving the full distribution of
the eigenvector components at the spectral edges, which have zero mean and are independent
of the spike prior.
These analytic results for the eigenvector distributions are
also relevant from the viewpoint of random matrix theory, since existing works characterize the leading eigenvector
of dense spiked random matrices mainly through its projection onto the spike vector \cite{B-GN,PottersNH}.

In the present model, both the outlier and the support of the continuous spectrum are
determined by the variance profile of the noise matrix. We solved our general equations for the case in which the variables defining this profile are sampled from a truncated
power-law distribution, controlled by a width parameter and a shape parameter. The resulting phase diagrams reveal
a rich non-monotonic behaviour of the BBP transition line as a function of the noise-inhomogeneity parameters.
In particular, this model exhibits a regime in which the critical signal-to-noise ratio decreases as the support
of the variance profile broadens, showing that noise inhomogeneity can, in some cases, favour signal detection.

From a methodological perspective, the present work combines the cavity method for dense
random matrices \cite{Cizeau1994} with an analytic approach to the statistics of the eigenvector
components outside the bulk, originally developed for sparse non-Hermitian
random matrices \cite{NeriMetz2016,MNR2019,Metz2021}. We have shown that this approach can also be
applied to symmetric matrices, with the simplification that no further
regularization of the resolvent is required in this case \cite{Feinberg1997}.
The general character of the present method allows one to study
leading eigenvector statistics and spectral inference in other symmetric spiked random-matrix models, including inhomogeneous
versions of the spiked Wishart ensemble \cite{BBP} and models with sparse noise \cite{Adomaityte25}. The present framework can also be extended to study
spectral inference in spiked non-Hermitian ensembles. Other important directions of future work include
  characterizing the information-theoretic limits of inference in the present  model, as well as investigating whether
  preprocessing strategies can exploit the noise structure to outperform unweighted PCA.
Overall, our results show that the structure of the noise matrix plays an important role in
high-dimensional inference, sometimes suppressing signal detection, and in other cases enhancing it.

\section*{Acknowledgements}

\paragraph{Funding information}

LSF acknowledges a fellowship from CNPq. FLM acknowledges financial support from CNPq (Grants No 402487/2023-0 and No 310255/2025-2) and from
ICTP through the Associates Program (2023-2028).


\begin{appendix}
\numberwithin{equation}{section}

\section{Derivation of the distributional equations}
\label{appecav}

In this appendix, we explain how Eqs. (\ref{juhrt}) and (\ref{evecs}) of the main text are obtained.
The derivation combines the cavity approach for fully-connected disordered systems \cite{Cizeau1994} with
identities originally derived in the context of non-Hermitian random matrices \cite{NeriMetz2016,MNR2019}.

We recall the definition of the $n \times n$ resolvent matrix $\boldsymbol{G}_n(z)=(z\boldsymbol{I}_n-\boldsymbol{A}_n)^{-1}$, with
$z=\lambda-i\epsilon$ and $\epsilon > 0$. We then introduce the normalized function of $\boldsymbol{\phi}_n$
\begin{equation}\label{complex-dist}
  P_z(\boldsymbol{\phi}_n|\boldsymbol{G}_n,\boldsymbol{\omega}_n) = \frac{1}{\mathcal{N}_n}
  e^{-\frac{i}{2}\boldsymbol{\phi}_n^T\boldsymbol{G}_n^{-1}\boldsymbol{\phi}_n+i\boldsymbol{\phi}_n^T \boldsymbol{\omega}_n},
\end{equation}
where $\boldsymbol{\omega}_n \in \mathbb{R}^n$ is an external field. The normalization constant is given by
\begin{equation}
    \mathcal{N}_n = \int_{\mathbb{R}^n} d\boldsymbol{\phi}_n e^{-\frac{i}{2}\boldsymbol{\phi}_n^T\boldsymbol{G}_n^{-1}\boldsymbol{\phi}_n + i\boldsymbol{\phi}^T_n \boldsymbol{\omega}_n},
\end{equation}
with $d\boldsymbol{\phi}_n = \prod_{i=1}^n d \phi_{i}$. The regularizer $\epsilon > 0$ ensures that integrals over the Gaussian
variables $\{ \phi_{i} \}_{i=1}^n$ are convergent.
Our initial goal is to represent the diagonal elements of $\boldsymbol{G}_n(z)$ and the eigenvector components
$R_{n,i}(\lambda)$ outside the bulk ($\lambda \notin \Omega_{\lambda}$) in terms of $P_z(\boldsymbol{\phi}_n|\boldsymbol{G}_n,\boldsymbol{\omega}_n)$, casting
the computation of these spectral properties into a statistical mechanics framework.
From Eqs. (\ref{res}) and (\ref{r-def}),  one can derive the identities \cite{NeriMetz2016,MNR2019}
\begin{eqnarray}\label{ap-res-1}
    &G_{n,ii}(z) = i\int_{\mathbb{R}^n} d\boldsymbol{\phi}_n\,\phi_{i}^2 P_z(\boldsymbol{\phi}_n|\boldsymbol{G}_n,\boldsymbol{\omega}_n=\boldsymbol{0}_n),\\
  \label{ap-res-2}
  & \left(\sum_{k=1}^n R_{n,k}  \right) R_{n,i}(\lambda) = -i \lim_{\epsilon \to 0^+} \epsilon \int_{\mathbb{R}^n} d\boldsymbol{\phi}_n \,\phi_{i}
    P_z(\boldsymbol{\phi}_n|\boldsymbol{G}_n,\boldsymbol{\omega}_n=\boldsymbol{1}_n),
\end{eqnarray}
where $\boldsymbol{0}_n=(0,\cdots,0)^T$ and $\boldsymbol{1}_n=(1,\cdots,1)^T$. Eqs. (\ref{ap-res-1})
and (\ref{ap-res-2}) express the spectral properties of interest in terms of the first two moments of
$P_z(\boldsymbol{\phi}_n|\boldsymbol{G}_n,\boldsymbol{\omega}_n)$.

Let us begin with the computation of the resolvent entries \cite{Cizeau1994}. We add a new site $i=0$ to the system, so that
the resolvent matrix gains an additional row and column comprising the interactions
between $i=0$ and the rest of the system. The diagonal element $G_{n+1,00}$ is then given by
\begin{eqnarray}
  \nonumber G_{n+1,00} &=& \frac{i}{\mathcal{N}_{n+1}} \int_{\mathbb{R}} d\phi_{0} \, \phi_{0}^2 \, e^{-\frac{i}{2}\phi_{0}^2 G_{n+1,00}^{-1}} \nonumber \\
  &\times&
  \int_{\mathbb{R}^{n}}\left(\prod_{i=1}^n d\phi_{i}\right) e^{-\frac{i}{2}\sum_{i,j= 1}^n\phi_{i}\phi_{j} G_{n,ij}^{-1} - i \phi_{0} \sum_{i=1}^n \phi_{i} G_{n+1,i0}^{-1}}.
\end{eqnarray}
Performing the Gaussian integration over $\{ \phi_i  \}_{i=1}^n$ yields
\begin{equation}
  G_{n+1,00} = i\frac{\int_\mathbb{R} d\phi_{0} \,\phi_{0}^2 \,
    e^{-\frac{i}{2}\phi_{0}^2\left(G_{n+1,00}^{-1}-\sum_{i,j= 1}^n G_{n+1,i0}^{-1}\, G_{n,ij}\, G_{n+1,0j}^{-1}\right)}}{\int_\mathbb{R} d\phi_{0} \,
    e^{-\frac{i}{2}\phi_{0}^2\left(G_{n+1,00}^{-1}-\sum_{i,j= 1}^n G_{n+1,i0}^{-1}\,G_{n,ij}\,G_{n+1,0j}^{-1}\right)}}.
\end{equation}
Substituting $\boldsymbol{G}_{n+1}^{-1} = z \boldsymbol{I}_{n+1} - \boldsymbol{A}_{n+1}$ and integrating over
$\phi_{0}$, we obtain
\begin{equation}\label{ap-res-3}
    G_{n+1,00}^{-1} = z-A_{n+1,00}-\sum_{i=1}^n A_{n+1,i0}^2 \, G_{n,ii} -\sum_{j\neq i =1}^n A_{n+1,i0}\,A_{n+1,0j}\,G_{n,ij},
\end{equation}
where $A_{n+1,ij}$ denotes the $ij$-th entry of $\boldsymbol{A}_{n+1}$. For large $n$, the off-diagonal
elements $\{ G_{n,ij} \}_{i \neq j =1}^n$ of the resolvent are of order $\mathcal{O}(1/\sqrt{n})$ \cite{Cizeau1994}.
Inserting the explicit form of $\boldsymbol{A}_{n+1}$, Eq. (\ref{model}), and applying the law of large numbers, we obtain the leading contribution to
$G_{n+1,00}^{-1}$,
\begin{equation}
G_{n+1,00}^{-1} = z - \frac{S_{n+1,0}}{n+1} \sum_{i=1}^n S_{n+1,i}  W_{n+1,i0}^2  G_{n,ii} + \mathcal{O} \left( \frac{1}{\sqrt{n}} \right).
\end{equation}  
In the limit $n \rightarrow \infty$, this expression yields the distributional equation (\ref{juhrt}) of the main text.

The computation of the eigenvector components follows along the same lines.
Adding a new site $i=0$ to the system and
integrating over $\{ \phi_i \}_{i=1}^n$ in Eq. (\ref{ap-res-2}), the eigenvector component $R_{n+1,0}$ reads
\begin{equation}
  \left(\sum_{k=0}^{n} R_{n+1,k}  \right) R_{n+1,0} = -i \lim_{\epsilon \to 0^+} \epsilon \, \frac{\int_\mathbb{R} d\phi_{0} \, \phi_{0} \,
    e^{-\frac{i}{2}  H_{n,0} \,  \phi_{0}^2 +i \phi_{0} \left( 1-\sum_{i,j=1}^n G_{n+1,i0}^{-1} G_{n,ij}  \right)  }}{\int_\mathbb{R} d\phi_{0} \,
    e^{-\frac{i}{2}  H_{n,0} \, \phi_{0}^2 +  i \phi_{0} \left(  1-\sum_{i,j=1}^n G_{n+1,i0}^{-1} G_{n,ij} \right)   }},
\end{equation}
where
\begin{equation}
H_{n,0} =  G_{n+1,00}^{-1}-\sum_{i,j=1}^n G_{n+1,i0}^{-1}\,G_{n,ij}\, G_{n+1,0j}^{-1}.
\end{equation}  
Computing the Gaussian integral over $\phi_0$ and substituting $\boldsymbol{G}_{n+1}^{-1} = z \boldsymbol{I}_{n+1} - \boldsymbol{A}_{n+1}$, we obtain
\begin{equation}
  \left(\sum_{k=0}^{n} R_{n+1,k}  \right) R_{n+1,0}
  = - i \lim_{\epsilon \to 0}  \frac{\epsilon \left( 1 + \sum_{i,j=1}^n A_{n+1,i0} G_{n,ij} \right) }{\left(z-A_{n+1,00}-\sum_{i,j=1}^n A_{n+1,i0}\,A_{n+1,0j}G_{n,ij}\right)}.
\end{equation}
Using Eq. (\ref{ap-res-3}), this expression can be rewritten as 
\begin{equation}
  \left(\sum_{k=0}^{n} R_{n+1,k}  \right)  R_{n+1,0} = -i \lim_{\epsilon \to 0} \epsilon \, G_{n+1,00}
  - i \lim_{\epsilon \to 0} \epsilon \, G_{n+1,00}  \sum_{i=1}^n A_{n+1,i0}  \sum_{j=1}^n G_{n,ij}.
\end{equation}
The first term on the right hand side vanishes as $\epsilon \rightarrow 0^{+}$, while the second term is related to the eigenvector
components in the original system with $n$ nodes through  Eq. (\ref{r-def}). This leads to the following
expression for finite $n$
\begin{equation}
   \left(\sum_{k=0}^{n} R_{n+1,k} \right) R_{n+1,0} =  \left(\sum_{k=1}^n R_{n,k}  \right) G_{n+1,00} \sum_{i=1}^n A_{n+1,i0} \,R_{n,i},
\end{equation}
where $G_{n+1,00}$ is the resolvent evaluated outside the bulk in the limit $\epsilon \rightarrow 0^{+}$. For large $n$, the sums over the eigenvector components on both sides
cancel, yielding the recursive equation
\begin{equation}
  R_{n+1,0} =   G_{n+1,00} \sum_{i=1}^n A_{n+1,i0} \,R_{n,i}.
\end{equation}
Substituting  Eq. (\ref{model}) in the above expression and then applying the central limit theorem and the law of large numbers, we recover
the distributional equation (\ref{evecs}) of the main text.

\end{appendix}
  
\bibliography{bib.bib}

\end{document}